\documentstyle[preprint,tighten,aps,prl]{revtex}
\begin{document}
\draft
\preprint{UTPT-94-16}
\title{Analysis of the Non-singular Wyman-Schwarzschild Metric}
\author{ Neil J. Cornish}
\address{Department of Physics, University of Toronto \\ Toronto,
Ontario M5S 1A7, Canada}
\maketitle
\begin{abstract}
The analog of the Schwarzschild metric is explored in the context of
Non-Singular Gravity. Analytic results are developed describing redshifts,
curvatures and topological features of the spacetime.
\end{abstract}
\narrowtext

\section{Introduction}
Recently it was discovered that a generalisation of Einstein's theory of
General Relativity (GR) yields an analog of the Schwarzschild
solution which is everywhere non-singular \cite{us,us2}. In these papers
it was established that the curvature invariants were finite throughout
the spacetime, and that no trapped surfaces occurred. The theory differs
from GR due to the presence of non-symmetric components in the metric. It is
the non-symmetric degrees of freedom which render the curvatures and
redshifts finite in this Non-Singular Gravity (NSG) theory.

The purpose of this paper is to explore more deeply the geometrical and
topological features of the Wyman-Schwarzschild spacetime in order to better
understand how singularities are avoided. The principle new insights are the
existence of a curvature plateau beginning at $r=2M$ and extending to the
origin, and a large increase in the non-Riemannian
character of the spacetime in this region. The departure from Riemannian
geometry is most pronounced near the origin where the effective dimension of
the spatial hypersurface approaches that of a two dimensional sphere.

The curvature plateau leads to
a roughly constant curvature region near the origin. A region of constant
curvature near the origin is reminiscent
of other non-singular analogues of the Schwarzschild metric which have been
studied in the literature\cite{them}. By introducing scalar fields or
{\it ad hoc} curvature cut-offs into a standard GR framework, the metrics in
Ref.\cite{them} were rendered non-singular by having a constant curvature
de Sitter-type region near the origin. Since infinite redshifts occur in
these metrics, the spacetimes are still singular in the strictest sense.
In contrast, the Wyman-Schwarzschild metric is truly non-singular since all
physical quantities remain finite.

In terms of a standard radial coordinate which defines circles of circumference
$2\pi r$, the proper volume is found to approach the usual value of
$4/3\, \pi r^3$ at large $r$, while near the origin the proper volume scales
as $r^2$. Since the volume scales as the surface area near the origin, we are
lead to the surprising conclusion that the spacetime is effectively
(2+1)-dimensional in that region. Moreover, the proper volume for the
Wyman-Schwarzschild metric is infinitely larger at $r=0$ than it is for the
Schwarzschild metric. It is this unique topological structure which renders
the curvature finite at the origin.

The transcendental form of the Wyman-Schwarzschild solution restricts our
analytic expansions to the regions $r/M<1$ and $2M/r >1$, so that numerical
results must be used to bridge the gap. These expansions also depend on
another dimensionless parameter, $s$, the so called skewness constant. The
skewness constant controls the departure of NSG from GR when $2M/r <\! < 1$.
For $r\leq 2M $, the limit $s \rightarrow 0$ is non-analytic, so that GR cannot
be recovered from NSG at or below the Schwarzschild radius. Typically, analytic
results can be found for $0<|s|<0.5$ and $|s|>1$, while numerical results must
again bridge the small gap in between.

\section{Field Equations}
Einstein's General Theory is based on the assumption that the metric
of spacetime is a symmetric tensor. By removing this assumption one enlarges
the group structure of the manifold's tangent space, thereby increasing the
number of gravitational degrees of freedom. This generalisation is due to
Moffat\cite{Moff79}, and has its roots in the old Einstein-Straus Unified
Field Theory. The Lagrangian for this theory is given by\cite{banff}
\begin{equation}\label{NGT}
{\cal L}=\sqrt{-g}g^{\mu\nu}\left(R_{\mu\nu}(\Gamma)+\frac{2}{3}W_{[\mu,\nu]}
-8\pi T_{\mu\nu}+\kappa W_{\mu}S_{\nu} \right)\; ,
\end{equation}
where $R_{\mu\nu}(\Gamma)$ is the generalized Ricci tensor, $W_{\mu}$ is
a Lagrange multiplier, $\Gamma$ refers to the torsion-free connection
$\Gamma^{\lambda}_{\mu\nu}$, $T_{\mu\nu}$ is the energy-momentum tensor
and $S_{\mu}$ is a conserved current. Square brackets denote
anti-symmetrisation
and units are chosen so that $G=c=1$ throughout.

The coupling to the current $S_{\mu}$ introduces equivalence-principle
violating effects into the theory. Strict adherence to the strong equivalence
principle is achieved by taking $\kappa=0$. With this choice, the field
equations that follow from (\ref{NGT}) are
\begin{eqnarray}
&& g_{\mu\nu,\sigma} - g_{\rho\nu} {\Gamma}^{\rho}_{\mu\sigma} -
g_{\mu\rho} {\Gamma}^{\rho}_{\sigma\nu} = 0 , \label{con} \\ \nonumber \\
&& {(\sqrt{-g}g^{[\mu \nu]})}_{ , \nu} = 0, \label{charge}
\\ \nonumber \\
&& R_{\mu \nu}(\Gamma) = \frac{2}{3} W_{[\nu , \mu]}+8\pi (T_{\mu\nu}
-\frac{1}{2}g_{\mu\nu}T) . \label{matter}
\end{eqnarray}
If we write the skew metric in terms of the Hodge decomposition:
\begin{equation}
\sqrt{-g}g^{[\mu \nu]}={\bf A}^{[\mu\nu]}+{\bf B}^{[\mu\nu]}\; ,
\end{equation}
where ${\bf A}$ and ${\bf B}$ satisfy
\begin{equation}
^{*}\! {\bf A}^{[\mu\nu]}_{\;\;\;\;\; ,\nu}=
\epsilon^{\mu\nu\kappa\lambda}{\bf A}_{[\kappa\lambda],\nu}=0 \; ,
\hspace{.3in} {\bf B}^{[\mu\nu]}_{\;\;\;\;\; ,\nu}=0 \; ,
\end{equation}
we find that (\ref{charge}) demands ${\bf A}^{[\mu\nu]}=0$. This restricts
$g^{[\mu\nu]}$ to the three ``axial'' components ${\bf B}^{[\mu \nu]}$. The
theory with $\kappa=0$ and ${\bf B}^{[\mu\nu]}\neq 0$ has been dubbed
Non-Singular Gravity (NSG) since it is known to have non-singular analogues of
the Schwarzschild and Reissner-Nordstrom spacetimes. General Relativity may be
thought of as coming from the above set of field equations, along with the
additional equation
\begin{equation}
{(\sqrt{-g}\, ^{*}\!g^{[\mu \nu]})}_{ , \nu} = 0, \label{GR}
\end{equation}
which demands ${\bf B}^{[\mu\nu]}=0$. When (\ref{GR}) is combined with
$(\ref{charge})$ we find $g^{[\mu\nu]}=0$, which enforces Einstein's original
assumption that the metric is symmetric.

\section{Wyman-Schwarzschild Metric}

The field equations (\ref{con})..(\ref{matter}) were first
solved for the static, spherically symmetric vacuum case by Wyman\cite{max}.
Wyman derived the solution for a mathematically identical set of equations
in Einstein-Straus Unified Field Theory. Because this unified theory failed
to describe nature, the Wyman solution was of little significance. It is
only when recast as a purely gravitational solution in the context of NSG
that the Wyman solution reveals its remarkable properties.

The metric can be written as:
\begin{equation}
ds^2=\gamma\, dt^2-\alpha \, dr^2-r^2 \, d\theta^2-r^2\sin^{2}\theta\, d\phi^2
\; ,\hspace{0.3in} g_{[\theta\phi]}=f\, \sin\theta \; . \label{met}
\end{equation}
The metric functions are given by
\begin{eqnarray}
\gamma&=&e^{\nu}\; , \\
\alpha&=&{M^2(\nu ')^2 e^{-\nu} (1+s^2) \over (\cosh(a\nu)-\cos(b\nu))^2}\; ,\\
\nonumber \\
f&=&{2M^2e^{-\nu}(\sinh(a\nu)\sin(b\nu)+s(1-\cosh(a\nu)\cos(b\nu)) \over
(\cosh(a\nu)-\cos(b\nu))^2} \; ,
\end{eqnarray}
where
\begin{equation}
a=\sqrt{{\sqrt{1+s^2}+1 \over 2}}\; , \hspace{0.5in}
b=\sqrt{{\sqrt{1+s^2}-1 \over 2}}\; ,
\end{equation}
prime denotes differentiation with respect to $r$, and $\nu$ is given
implicitly
by the relation:
\begin{equation}\label{impl}
e^{\nu}(\cosh(a\nu)-\cos(b\nu))^2{r^2 \over 2M^2}=\cosh(a\nu)\cos(b\nu)
-1+s\sinh(a\nu)\sin(b\nu) \; .
\end{equation}
The Schwarzschild solution is recovered by taking the limit $b\nu \rightarrow
0$. At some large distance $r$ from the
origin, $\nu$ is a small constant and the limit $b\nu \rightarrow 0$ is
equivalent to the limit $s \rightarrow 0$. Near the origin $\nu$ becomes
large and the product $b\nu$ cannot be made to vanish, as will be shown below.

For $r>2M$ the solution rapidly approaches the Schwarzschild metric and
takes the explicit form:
\begin{eqnarray}
\gamma&=&1-{2M \over r}+{s^2 M^5 \over 15 r^5}+{4 s^2 M^6 \over 15 r^6}
        + \dots \; , \label{gs} \\ \nonumber \\
\alpha&=&\left(1-{2M \over r}+{2s^2M^4 \over 9r^4}
+{7s^2M^5 \over 9r^5}+{87s^2M^6 \over 45 r^6} + \dots \right)^{-1}
\; , \\ \nonumber \\
f&=&{sM^2 \over 3}+{2sM^3 \over 3r}+{6sM^4 \over 5r^2} + \dots \; .\label{fs}
\end{eqnarray}
Near $r=0$ the Wyman metric takes on the very different form:
\begin{eqnarray}
\gamma&=&\gamma_{0} +{\gamma_{0}(1+{\cal O}(s^2)) \over 2|s|}
\left({r \over M}\right)^2+{\cal O}((r/M)^4)\; , \label{g}\\ \nonumber \\
\alpha&=&{4\gamma_{0}(1+{\cal O}(s^2)) \over s^2}\left({r \over M}\right)^2
+{2\gamma_{0}(1+{\cal O}(s^2)) \over s^3}\left({r \over M}\right)^4
+{\cal O}((r/M)^6)\; , \label{a} \\ \nonumber \\
f&=&M^2\left(4-{|s|\pi \over 2}+s|s|+{\cal O}(s^3)\right)
+{|s|+s^2\pi/8+{\cal O}(s^3) \over 4}r^2+{\cal O}(r^4)\; ,\label{f}
\end{eqnarray}
where $\gamma_{0}$ is given by
\begin{equation}
\gamma_{0}=\exp{\left(-{\pi +2s \over |s|}-{s \pi  \over 8}
+{\cal O}(s^2)\right)} \; .
\end{equation}
The above expansion was for $|s|<1$, but expansions with $|s|>1$ can also
be found. The expansions with $|s|>1$ retain the same form as a series in
$r/M$, but the coefficients change. For example, $\gamma_{0}$ becomes
\begin{equation}
\gamma_{0}=\exp\left(-\pi\sqrt{{2 \over |s|}}\left[1-{\pi|s|-s(1+2e^{-\pi})
+{\cal O}(s^0) \over 2\pi s^2}\right]\right) \; .
\end{equation}
The most important feature of the solution can be seen from (\ref{f}), where
it is clear that the limit $|s|\rightarrow 0$ does not recover the GR metric
with $f=0$. The product $b\nu$ approaches $-\pi /2$ and cannot be made to
vanish as $|s| \rightarrow 0$.
The only way to have $f=0$ globally is to put $M=0$ whereby both the
GR and NSG solutions reduce to Minkowski space. Thus, the limit
$|s|\rightarrow 0$ is non-analytic for $2M/r\geq 1$ and analytic for $2M/r <1$.

\subsection{Uniqueness of the Wyman-Schwarzschild Solution}

In General Relativity, Birkhoff's theorem guarantees that a spherically
symmetric gravitational field in empty space must be stationary, with a
metric given by the Schwarzschild metric (only the region outside of $r=2M$
is the metric actually static). No direct generalization of Birkhoff's theorem
exist for NSG. What can be shown is the following. (I) The Wymann-Schwarzschild
solution is the unique, static spherically symmetric spacetime which is
asymptotically Minkowskian. (II) There is no monopole radiation in any
spherically symmetric spacetime which is asymptotically Minkowskian. (III) If
the skew field falls off faster than $1/r$ in orthonormal coordinates, a
spherically symmetric gravitational field in empty space must be static, with a
metric given by the Wyman-Schwarzschild metric, i.e. a generalised Birkhoff
theorem then holds.

The essential difference between GR and NSG for the spherical case is that
NSG admits solutions which are not asymptotically flat. In that sense, the
Wyman-Schwarzschild solution, as described in the previous section, is not
unique. This is analogous to the situation in GR where the Kerr solution is
not the unique axisymmetric vacuum solution, although it is the unique
asymptotically Minkowskian axisymmetric vacuum solution. Points (I), (II) and
(III) can most easily be proved as follows. Consider the most general,
asymptotically Minkowskian, spherically symmetric metric in NSG, which may be
written as an expansion in inverse powers of $r$ for large $r$:
\begin{eqnarray}
\alpha(r,t)&=&1+{\alpha_{1}(t) \over r}+{\alpha_{2}(t) \over r^2}+\dots \; ,
\\ \nonumber \\
\gamma(r,t)&=&1+{\gamma_{1}(t) \over r}+{\gamma_{2}(t) \over r^2}+\dots \; ,
\\ \nonumber \\
f(r,t)&=&f_{1}(t)r+f_{2}(t)+{f_{3}(t) \over r}+\dots \; .
\end{eqnarray}
The coefficient $f_{1}(t)$ is the coefficient of the $1/r$ term of $f$ in
orthonormal coordinates.

To leading order, the field equations give
\begin{equation}
{\partial \alpha_{1} \over \partial t}={\partial \alpha_{2} \over \partial t}=
{\partial \gamma_{1} \over \partial t}={\partial \gamma_{2} \over \partial t}=
{\partial f_{1} \over \partial t}= {\partial^2 f_{2} \over \partial t^2}=0\; ,
\end{equation}
and
\begin{eqnarray}
&& \alpha_{1}=-\gamma_{1}=2M \; , \\ \nonumber \\
&& {\partial f_{2} \over \partial t}={\sqrt{7} \over 2}f_{1} \; , \label{st}
\\ \nonumber \\
&& \gamma_{2}={5 \over 8}f_{1}^2 \; , \\ \nonumber \\
&& \alpha_{2}=M^2-{27\over 16}f_{1}^2 \; .
\end{eqnarray}
Continuing to higher orders yields the same expansions as in (\ref{gs})..(
\ref{fs}), along with many new terms, all of which are proportional to $f_{1}$.
If the spacetime is static, (\ref{st}) demands $f_{1}=0$ and the solution
collapses back to the Wyman-Schwarzschild metric, thus proving point (I).
Similarly, if we demand that $f$ fall off faster than $1/r$, $f_{1}$ is
again zero, thus proving point (III). Even if $f_{1}\neq 0$, we still have
$\partial_{t} \alpha_{1}=2 \partial_{t} M=0$, so there is no monopole
radiation, which proves point (II).

\section{Geometry and Topology}

The deviation from Riemannian geometry is controlled
by the size of the invariant skew potential
\begin{equation}
F^{2}=g_{[\mu\nu]}g^{[\mu\nu]}={f^2 \over r^4 +f^2} \; .
\end{equation}
At large $r/M$ this quantity approaches
\begin{equation}
F^2={s^2M^4 \over 9r^4}\left(1+{4M\over r}+\dots\right) \; ,
\end{equation}
while for $r<2M$ the skew invariant is given to leading order in $s$ by
\begin{equation}
F^2=1-\left(1+{|s|\pi \over 4}\right)\left({r \over 2M}\right)^4+{|s| \over 2}
\left({r \over 2M}\right)^6+\dots \; .
\end{equation}
We see that $F$ is of order one below the Schwarzschild radius and
essentially zero above $r=2M$, with rapid $1/r^2$ fall-off
at spatial infinity. The geometry of spacetime at and below
$r=2M$ is dominantly non-Riemannian. This is evidenced by the form of
geometrical
\
\begin{figure}[h]
\vspace{60mm}
\includegraphics{logV.ps}
\vspace{20mm}
\caption{$\log_{10}(VM^{-3})$ for NSG with $s=0.1$
(solid line) and GR (dashed line).}
\end{figure}
\noindent quantities such as the curvature and proper volume. The
proper volume of a region with radius $r$ is given by
\begin{eqnarray}
V&=& 4\pi \int_{0}^{r}\sqrt{-g}\, dr\; , \nonumber \\ \nonumber \\
&=&8\pi M^3 (1+s^2) \int_{\nu(0)}^{\nu(r)} { e^{-\nu} d\nu \over
(\cosh(a\nu)-\cos(b\nu))^2} \; .
\end{eqnarray}
For large $r$ this expression yields the standard result $4/3 \pi r^3$ while
for small $r/M$ and small $|s|$ the proper volume is given by
\begin{equation}
V=4\pi M r^2\, {4-\pi|s|/2+s^2+\dots \over |s|}\exp{\left(-{\pi +2s \over |s|}
-{s \pi  \over 8}\right)}\left[1+{1\over s}\left({r \over 2M}
\right)^2+{2 \over 3 s^2}\left({r \over 2M}\right)^4+\dots \right] \; .
\end{equation}
In the limit $r/M \rightarrow 0$ the volume scales as $r^2$ rather than
$r^3$, making the volume infinitely larger than in GR. For $r/M > 12\exp(
-\pi/|s|)/|s|$ the situation is reversed with GR having a larger proper
volume than NSG. A logarithmic plot of the proper volume is displayed in
Figure 1.

A useful measure of the non-Riemannian nature of the geometry is
the ``effective dimension'' of spacetime. While the physical dimension
remains fixed at four, the volume of appropriately chosen spatial
hypersurfaces may scale quite differently. Consider a locally inertial
coordinate system in the neighbourhood of some point so that
$g_{(\mu\nu)}=\eta_{\mu\nu}$ and $g_{(\mu\nu),\sigma}=0$. Adopting
spherical coordinates about this point, the volume of a infinitesimal
sphere of radius $r$ is given by
\begin{equation}
V \propto r^d \; ,
\end{equation}
where $d$ is the effective dimension of the spatial section. A useful
expression for $d$ is
\begin{equation}
d=1+r{V'' \over V'}=1+{rg' \over 2g} \; .
\end{equation}
In GR, coordinates can always be found in the neighbourhood of a point
such that $g=-r^4\sin^{2}\theta$ so that $d=3$. In contrast, the
diffeomorphism invariance of NSG is only able to make the symmetric
metric Minkowskian so that $d\neq 3$ unless $g_{[\mu\nu]} =0$. Of
course, the dimension is very close to three when the skew invariant
$F$ is small.

For a spherically symmetric spacetime in GR it is always possible
(locally) to choose a radial coordinate which foliates two-surfaces
of area $4\pi r^2$ such that the proper volume between surfaces at $r$ and
$r+\delta r$ is $4\pi r^2 \delta r$. Again, this is not possible in
NSG because $F$ cannot be removed by a coordinate transformation. To
faithfully study the departure of NSG from Riemannian geometry would
require a local redefinition of the radial coordinate so that the
symmetric metric is such that $(g_{s}={\rm det}g_{(\mu\nu)})$
\begin{equation}
d_{s}=1+{r g_{s}' \over 2g_{s}} =3 \; .
\end{equation}
The total effective dimension for the Wyman-Schwarzschild metric can then
be written as
\begin{equation}
d=1+{r g_{s}' \over 2g_{s}}+{rFF' \over 1-F^2}=3+{rFF' \over 1-F^2} \; .
\end{equation}
In terms of the global radial coordinate we chose earlier, $F' <0$.
Any new radial coordinate ${\bar r}$ which satisfies
$\partial {\bar r} / \partial r >0$ will have $F' <0$ also so that the
effective dimension will be $\leq 3$.
\
\begin{figure}[h]
\vspace{60mm}
\includegraphics{Dim.ps}
\vspace{25mm}
\caption{The dimension $d$ for NSG with $s=0.1$}
\end{figure}

A cruder, but much simpler effective dimension to study results from choosing
a global coordinate system with a radial variable which foliates
two-surfaces of area $4\pi r^2$ and calculating $d(r)$ for this coordinate
choice. For the Schwarzschild metric this gives $d(r)=3$, although this
result is only meaningful for $r>2M$ where the coordinate patch is valid. For
the Wyman-Schwarzschild metric the analogous coordinate system is globally
valid and gives $d=3$ for $r\rightarrow \infty$ and $d=2$ as $r
\rightarrow 0$. A plot of this effective dimension is displayed in
Figure 2. Since this coordinate system does not ensure $d_{s}=3$, we
find that $d>3$ in some regions. This can be traced to the fact that
$(\log(\alpha\gamma))'$ gets very large beneath $r=2M$ for this choice of
radial variable.

Although each of the effective dimensions described above have different
properties, they all serve to measure the departure
from Riemannian geometry. The effective dimension defined for
an infinitesimal sphere in locally inertial coordinates is perhaps the
closest to a true topological measure, while the dimension displayed in
Figure 2. has the advantage of being easily calculated.

\section{Redshifts}

The components of the Killing vectors, $^{(K)}\xi^{\mu}$, for the
non-Riemannian geometry of NSG are given by
\begin{equation}
g_{\mu\sigma} \, ^{(K)}\xi^{\sigma}_{\; ,\nu}
+g_{\sigma\nu}\, ^{(K)}\xi^{\sigma}_{\; ,\mu}
+g_{\mu\nu,\sigma}\, ^{(K)}\xi^{\sigma}=0 \; .
\end{equation}
For the metric (\ref{met}) the Killing vector which is timelike at spatial
infinity is given by
\begin{equation}
^{(t)}\xi_{\mu}=(\gamma,0,0,0) \; .
\end{equation}
The norm of this Killing vector,
\begin{equation}
^{(t)}\xi^{\mu}\, ^{(t)}\xi_{\mu}=\gamma \; ,
\end{equation}
never vanishes because $\gamma>0$ throughout the spacetime. Figure 3. displays
$\gamma$ for NSG with the choice $s=1$ and for GR where the norm of the
timelike Killing vector, $g_{tt}\equiv \gamma^{{\rm GR}}$, vanishes at $r=2M$.
So long as $s \neq 0$ there are no null surfaces in the spacetime and no
Hawking radiation.

For $s \neq 0$ the Wyman solution is characterised by finite redshifts
between any two points in the spacetime. The maximum redshift occurs between
$r=0$ and $r=\infty$. This redshift can be calculated analytically in terms
of both small and large $s$ expansions. The maximum redshift is given by
\begin{equation}
z_{max}={1 \over \sqrt{\gamma_{0}} }-1\; .
\end{equation}
If the skewness constant $s$ is small, $z_{max}$ is given by
\begin{equation}
z_{max}=\exp\left({\pi+2s+ s^2\pi/8  +{\cal O}(s^3) \over 2|s|}\right) -1\; ,
\end{equation}
while if $s$ is large, $z_{\max}$ is given by
\begin{equation}
z_{max}=\exp\left({\pi \over \sqrt{2|s|}}\left[1-{\pi|s|-s(1+2e^{-\pi})
+{\cal O}(s^0) \over 2\pi s^2}\right]\right)-1 \; ,
\end{equation}
\begin{figure}[h]
\vspace{70mm}
\includegraphics{gam1.ps}
\vspace{20mm}
\caption{The norm of the timelike Killing vector for NSG with $s=1$
(solid line) and GR (dashed line).}
\end{figure}
Both of these expansions give excellent approximations to the exact numerical
results in their respective regions of validity. The large $s$ expansion is
very accurate in the range $1<|s|<\infty$, and the small $s$ expansion is
very good in the range $0<|s|<0.5$. Even in the intermediate region $0.5<|s|<1$
both expansions are good to within $\sim 10\%$. The utility of these
expressions
is made clear in Figure 4.

Similar results can be obtained for the redshift between surfaces with
$r<2M$ and $r=\infty$. For example, the redshift between $r=\sqrt{2}\, M$
and $r=\infty$ is given by
\begin{equation}
z_{{\sqrt{2}}}=\exp\left({\pi \over 3|s|}+{\cal O}(s^{0})\right)-1 \; ,
\end{equation}
for small $|s|$. The transcendental relation for $\gamma$ does not lend
itself to a simple inversion in the immediate neighbourhood of $r=2M$, although
the following simplified transcendental relationship can be found for
$\gamma_{2M}$ when $|s|<<1$:
\begin{equation}
{e^{\pi} \over \gamma_{2M}}\log\left({e^{\pi} \over \gamma_{2M}}\right)=
{16 \over s^2}+{\cal O}(s^{-1})\; .
\end{equation}
The above relation shows that as $|s|$ tends to zero $\gamma_{2M}$ tends to
zero, leading to a surface of infinite redshift. However, so long as $|s|$ is
non-zero the redshift remains finite everywhere. The redshift between
$r=2M$ and $r=\infty$ as a function of $s$ is plotted in Figure 5.
\newpage

\
\begin{figure}[h]
\vspace{70mm}
\includegraphics{logred.ps}
\vspace{25mm}
\caption{The maximum redshift is plotted as a function of $s$ on a logarithmic
scale. The circles are the exact, numerical result, while the dashed line is
the analytic expression for small $s$ and the solid line is the analytic
expression for large $s$.}
\end{figure}

\
\begin{figure}[h]
\vspace{60mm}
\includegraphics{logred2m.ps}
\vspace{20mm}
\caption{The logarithm of the redshift from $r=2M$ to $r=\infty$ is plotted
as a function of $s$. For small $s$ the redshift becomes very large, but
remains
finite so long as $s\neq 0$.}
\end{figure}

\section{Curvature}
In order to decide whether a geometrical gravity theory is non-singular
the curvature invariants must be computed. If the curvature invariants
are finite, the tidal forces on a body will be finite also.

In the weak field region where $2M/r <1$, the curvature is very similar
to what one finds in GR. For example, the generalised Kretschmann scalar
is given by
\begin{eqnarray}
K&=&R^{\mu\nu\kappa\lambda}R_{\mu\nu\kappa\lambda} \\
&=&{48 M^2 \over r^6}\left(1+{5 s^2 M^2 \over 12 r^2}+
{s^2 M^3 \over r^3}+{55 s^2 M^4 \over 27 r^4}+\dots \right) \; .
\end{eqnarray}
The other invariants have similar forms such as
\begin{equation}
R^{\mu\nu}_{\;\;\; \kappa\lambda}R^{\kappa\lambda\sigma\rho}R_{\sigma\rho
\mu\nu}={96 M^3 \over r^9}\left(1+{s^2 M^3 \over 6r^3}+{5 s^2 M^4 \over 18 r^4}
+\dots \right) \; .
\end{equation}
At large values of $r/M$, the curvature in NSG is greater than in GR, while
for $r/M \rightarrow 0$ the opposite is true.
The large $r$ expansion is valid up to, but not including, $r=2M$. The series
are very slow to converge near $2M$. Near $r=0$ all the non-vanishing
curvatures scale as $1/M^2$ and are approximately constant for $r/M < |s|$
when $0<|s|<0.5$. As an example, the Kretschmann scalar is given by
\begin{equation}
K={s^4(1+|s|\pi /2+\dots) \over 2^9 \, M^4}\exp\left({2\pi+4s+s^2 \pi /4 +\dots
\over |s|}\right)\left[1-{(1+\dots) \over |s|}{r^2 \over M^2}-\dots\right] \; .
\end{equation}

\begin{figure}[h]
\vspace{70mm}
\includegraphics{logK.ps}
\vspace{20mm}
\caption{The scaled Kretschmann scalar, $\log_{10}(KM^4 +1)$, plotted as
a function of $r/M$ for both GR (dashed line) and NSG (solid line).}
\end{figure}
The Kretschmann scalar is displayed in Figure 6. for both NSG with
$s=0.1$ and GR. The quantity being plotted is actually the dimensionless
quantity $\log_{10}(KM^4+1)$ for ease of display.

Unlike modifications to GR which are designed to be important for large
curvatures\cite{them}, the curvature in NSG differs enormously from GR even
where GR has small curvatures. Moreover, the curvature in NSG never has to be
large since it is proportional to $1/M^2$ for all $r/M$. For
large enough masses the curvature is small everywhere. For this reason it
is misleading to call NSG a strong field modification of GR. NSG differs from
GR not only in regions where GR predicts large curvatures, but also in small
curvature regions such as a Schwarzschild radius from the centre of a
$10^8$ solar mass object. A more accurate statement is that NSG is a
{\em large redshift} modification of GR. Regions of high redshift are a
natural place to expect a high-energy modification of GR to be important,
whether that modification be an alternative classical theory or
quantum GR. Large redshifts dilate
time and magnify space allowing the high-energy physical processes going on
in these regions to become accessible to the effective low-energy physics
outside. Perhaps for NSG we can say that this magnifying-glass effect reveals
the underlying non-Riemannian character of the spacetime geometry which is
otherwise unimportant on standard time and length scales.

\section*{Acknowledgments}
I am grateful for the support provided by a Canadian Commonwealth
Scholarship. I thank Mike Clayton, Janna Levin, John Moffat, Igor Sokolov
and Glenn Starkman for many interesting discussions concerning NSG.

\end{document}